%% file: arxiv.tex
\documentclass[preprint]{vgtc}

\graphicspath{{figure/}{./}}

\usepackage{times}

\usepackage{mathptmx}

\input{preamble_arxiv}

\preprinttext{}
\nocopyrightspace

\vgtcinsertpkg

\title{ReRoom: Blending Virtual and Physical Contexts for \\
In Situ Room Planning in Mixed Reality}

\input{arxiv_authors}

\teaser{
  \centering
  \includegraphics[width=\linewidth]{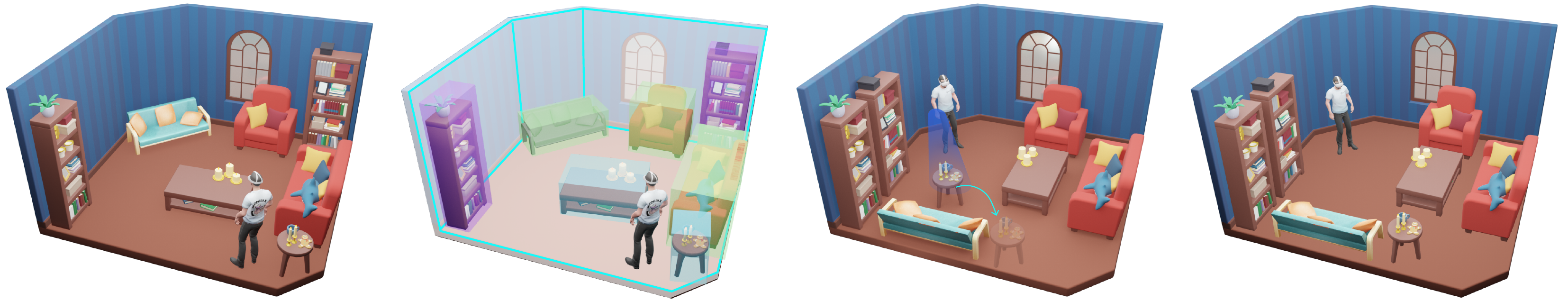}
  \caption{\SysName{} supports in situ iterative room planning in mixed reality. From left to right, a user starts in an existing room, scans the space with an MR headset to reconstruct the room geometry (blue wireframe) and semantically labeled furniture, receives a new layout proposal from \SysName{}'s skill-guided layout agent, and then refines the result through direct manipulation (e.g., adjusting a table position) before obtaining the final room arrangement.}
  \label{fig:teaser}
}

\abstract{

\input{text/0_abstract}
}


\hypersetup{
  pdftitle={ReRoom: Blending Virtual and Physical Contexts for In Situ Room Planning in Mixed Reality},
  pdfauthor={\ArxivPdfAuthors},
  pdfkeywords={Room Layout Planning, Mixed Reality, In Situ Interaction, LLM-based agents}
}

\begin{document}

\maketitle

\input{text/1_introduction}
\input{text/2_related}
\input{text/3_method}
\input{text/4_implementation}
\input{text/5_technical_eval}
\input{text/6_user_study}

\input{text/7_conclusion}

\bibliographystyle{abbrv-doi}
\bibliography{refs}

\clearpage
\appendix
\section*{Supplementary Materials}
\setcounter{section}{0}
\setcounter{figure}{0}
\setcounter{table}{0}
\setcounter{equation}{0}
\renewcommand{\thesection}{S\arabic{section}}
\renewcommand{\thefigure}{S\arabic{figure}}
\renewcommand{\thetable}{S\arabic{table}}
\renewcommand{\theequation}{S\arabic{equation}}
\renewcommand{\theHsection}{supp.\arabic{section}}
\renewcommand{\theHfigure}{supp.\arabic{figure}}
\renewcommand{\theHtable}{supp.\arabic{table}}
\renewcommand{\theHequation}{supp.\arabic{equation}}

\input{text/supp}

\end{document}

%% file: preamble_arxiv.tex
\usepackage[svgnames]{xcolor}
\usepackage{amsmath}
\usepackage{booktabs}
\usepackage{xurl}
\usepackage{tabularx}
\usepackage{multirow}
\usepackage{enumitem}
\usepackage{tcolorbox}
\tcbuselibrary{listings,breakable}

\newcommand{\SysName}[0]{\textbf{ReRoom}}
\newcommand{\BaselineName}[0]{\textbf{Baseline}}

\definecolor{UserStudyBaseline}{HTML}{6CB4EE}
\definecolor{UserStudyReRoom}{HTML}{F5C78E}
\definecolor{PerceptGLTScene}{HTML}{7FB3D8}
\definecolor{PerceptLayoutVLM}{HTML}{A3D977}
\definecolor{PerceptReRoom}{HTML}{F5C78E}

%% file: arxiv_authors.tex
\newcommand{\ArxivPdfAuthors}{Hongliang Yang, Yanjing Xu, Anhang Zhang, Hui Ye, and Pengfei Xu}
\author{Hongliang Yang\textsuperscript{1} \quad Yanjing Xu\textsuperscript{1} \quad Anhang Zhang\textsuperscript{1} \quad Hui Ye\textsuperscript{2} \quad Pengfei Xu\textsuperscript{1,*}}
\affiliation{\scriptsize
\textsuperscript{1}College of Computer Science and Software Engineering, Shenzhen University, Shenzhen, Guangdong, China\\
\textsuperscript{2}Department of Interactive Media, Hong Kong Baptist University, Hong Kong, China\\
\textsuperscript{*}Corresponding author: \href{mailto:xupengfei.cg@gmail.com}{xupengfei.cg@gmail.com}}

%% file: text/0_abstract.tex
Planning a real domestic space is an in situ authoring process: users evaluate candidate layouts at true scale, refine their intent, and carry accepted decisions into later iterations. Existing approaches either separate layout editing from the physical room or provide limited support for evaluating and refining whole-room proposals in situ. We present \SysName{}, a mixed-reality system for in situ room-layout authoring. \SysName{} presents a shared layout state through a virtual room proxy spatially registered to the target room, allowing interaction and layout generation to remain grounded in the physical context. Users refine the current proposal through direct manipulation or language and preserve accepted placements, allowing each generated update to continue the same evolving design. To balance layout quality with generation efficiency, \SysName{} uses a skill-guided layout agent whose room-layout design skill operationalizes three principles that we formulate by synthesizing established interior-design guidance for real-room layout generation. The skill grounds these principles in a normalized representation of the scanned room and reusable geometric checks. Evaluations show that \SysName{} produces high-quality layouts for non-rectangular rooms, while its in situ workflow improves the room-planning experience over an otherwise equivalent off-site VR workflow. Code will be released upon acceptance of the paper.

%% file: text/1_introduction.tex
\section{Introduction}

Planning a real domestic space requires users to balance functional needs, circulation, spatial dimensions, and personal preferences as the design considerations. Unlike one-shot generation, users rarely begin with a complete and precise description of an ideal layout. Many implicit constraints become apparent only when a candidate layout is presented at a true scale within the target room, where it can be judged against the walls, doorways, existing furniture, and lighting conditions. In-place observation may reveal, for example, that a dining table obstructs the entrance, a sofa blocks an important sightline, or an already satisfactory area should remain unchanged in later iterations. Room-layout planning is therefore fundamentally a form of in situ authoring: after the system proposes a candidate layout, users draw on the room's physical scale and context to assess it, revise local arrangements in response to problems discovered on site, and preserve accepted design decisions. Each revised layout serves as the starting point for the next round, allowing users to clarify their design intent and refine the layout proposal until they are satisfied.

Although existing tools support some stages of this process, none supports the full process within the target physical room. Conventional 2D floor-plan and desktop 3D tools avoid the labor of moving physical furniture, but force users to perform cognitively demanding mental mappings between a flat display and the real room, making it difficult to accurately perceive crowding, sightline occlusion, and the overall spatial atmosphere at true scale. Mobile AR applications such as IKEA Place~\cite{ikea2017place} and Apple RoomPlan~\cite{apple2022roomplan} overlay virtual furniture on the physical environment, but their limited field of view and item-by-item placement model do not support systematic evaluation of a whole-room layout. Immersive VR systems provide natural 3D editing~\cite{zhang2024vrcopilot,hou2025echoladder}, but visually isolate users from the physical environment, preventing in situ comparison of candidate layouts with the room's actual walls, doors, windows, lighting, and existing facilities. Even in a carefully scale-matched VR replica, size judgments can differ significantly from those made during real-world viewing~\cite{wijayanto2023size}. Beyond these interaction limitations, complete in situ authoring also places distinct demands on the underlying layout-generation method. Most existing automated layout-generation methods primarily produce standalone layouts from text descriptions or abstract constraints~\cite{li2019grains,paschalidou2021atiss,para2023cofs,sun2025layoutvlm,li2024gltscene}, rather than update an evolving design in place. In situ authoring instead requires each update to respond to new intent while preserving accepted decisions and satisfying the scanned room's physical constraints. Together, these requirements call for a room-planning system that integrates in situ authoring with high quality layout generation for scanned, irregular rooms.

To meet these requirements, we present \SysName{}, a mixed-reality system for authoring real-room layouts in situ. \SysName{} maintains the evolving proposal in a shared layout state that connects all authoring operations, allowing edits and accepted decisions to persist across planning rounds. A virtual room proxy renders this state as a candidate layout spatially registered to the target room, returning every update to the room's true scale and physical context for inspection. \SysName{} also uses a skill-guided layout agent to generate layout updates for scanned physical rooms. The virtual room proxy and layout agent remain synchronized through the shared state.

In a typical workflow, the user requests an initial candidate layout through natural language, and \SysName{} presents it at real scale in the virtual room proxy spatially registered to the target room. By adjusting the opacity between the passthrough view and the proxy, users compare the existing room with the proposal, revealing needs that were difficult to anticipate or describe in advance. Users then translate or rotate furniture through direct manipulation for precise local changes, or express higher-level intent through language-guided editing. Once satisfied with an item or group, they can apply intent locking to preserve its placement; in later generation requests, the layout agent retains these decisions and reorganizes the unlocked portion according to the design principles encoded in its domain skill. The updated layout is returned to the proxy for another round of in situ observation and refinement.

Because each revision in this process invokes the layout backend before the updated proposal can be inspected in situ, the workflow places coupled demands on layout quality and generation efficiency. Each update should satisfy the room's spatial and functional requirements while being returned quickly enough to sustain the observation–refinement loop.

To balance these objectives, we draw on established interior-design guidance~\cite{kilmer2014designing,panero1979human} to formulate three principles for agent-based real-room layout generation: functional grouping, anchor furniture selection, and geometry-guided refinement. We encode these principles in a room-layout design skill and pair the skill with an axis-normalized spatial representation and reusable geometric checks. This design structures the agent's spatial reasoning and focuses its computation on proposing and revising layouts, improving generation efficiency while preserving its ability to adapt to different rooms and user requests.

We evaluate both the quality of layouts generated by \SysName{} and the in situ authoring experience supported by the complete system. A quantitative evaluation on 43 rooms, nearly all with non-rectangular boundaries, compares the skill-guided layout agent with GLTScene~\cite{li2024gltscene} and LayoutVLM~\cite{sun2025layoutvlm} across six physical and semantic layout-quality metrics; \SysName{} achieves the best result on all six. In a perceptual evaluation, 45 independent raters gave \SysName{}'s layouts significantly higher ratings than both baselines for physical plausibility, functional coherence, and overall preference. In a 14-participant user study, we compare the complete \SysName{} system with an off-site VR workflow that provides the same interaction functions and layout-generation capability but no physical-room reference. Relative to the off-site VR workflow, in situ authoring improves spatial perception and confidence in layout decisions, reduces overall task load, and increases system usability.

Our contributions are:
\begin{itemize}[leftmargin=10pt, nosep]
    \item We present an in situ room-layout authoring workflow that presents a shared layout state through a virtual room proxy spatially registered to the target room, allowing all authoring operations to remain grounded in the physical context.
    \item We introduce a room-layout design skill that structures general-purpose LLM reasoning for real-room layout generation by operationalizing three design principles we formulate from established interior-design guidance.
    \item We evaluate the quality of \SysName{}'s generated layouts. A user study shows that authoring these layouts in the target physical room improves the room-planning experience.
\end{itemize}

%% file: text/2_related.tex
\section{Related Work}

\subsection{Immersive Spatial Authoring and Editing}
Immersive authoring systems address one side of our problem by making spatial editing more direct and interactive. We organize recent work along a continuum from fully virtual to mixed-reality environments. 

Recent virtual-reality systems bring generative AI into the authoring loop to improve user control~\cite{tsai2026generative, yin2024text2vrscene, chen2025multimodal, zhang2024vrcopilot, hou2025echoladder, jiang2024vr}. VRCopilot~\cite{zhang2024vrcopilot} introduces wireframes as low-fidelity intermediate representations and supports manual, scaffolded, and automatic modes of human-AI co-creation, finding that scaffolded wireframe editing significantly increases user agency over fully automatic generation. EchoLadder~\cite{hou2025echoladder} uses a large vision-language model to translate verbal instructions at varying abstraction levels into modular, interactable design suggestions that users can selectively apply, undo, or regenerate, thereby externalizing the AI reasoning process for finer-grained user control. Meanwhile, VR-GS~\cite{jiang2024vr} enables physics-aware world editing within reconstructed scenes. These systems demonstrate the value of AI-assisted immersive authoring, yet they operate in reconstructed or fully virtual scenes, which limits in situ evaluation relative to the user's actual room.

Moving toward physical grounding, mixed-reality authoring systems~\cite{yue2017scenectrl,wang2023pointshopar,yang2025mrpilot,kari2023scene,zhao2025guided} enable users to interact with and edit scenes within their real environment. SceneCtrl~\cite{yue2017scenectrl}, PointShopAR~\cite{wang2023pointshopar}, MRPilot~\cite{yang2025mrpilot}, and Scene Response~\cite{kari2023scene} support operations such as object removal, point cloud editing, task navigation, and just-in-time virtualization of physical objects, thereby reducing the friction of interacting with one's space in situ. Reality Promises~\cite{kari2025reality} pushes this further by coupling MR perception with physical actuation: an invisible mobile robot secretly replicates virtual object manipulations in the real world, using robot-aware 3D Gaussian splatting to hide the robot from the user's passthrough view. While it demonstrates that MR systems can bridge the gap between virtual edits and physical outcomes, its focus is on illusory single-object interactions rather than layout planning. LLMR~\cite{de2024llmr} introduces language-driven scene creation that additively places generated objects into the physical environment. DreamSpace~\cite{yang2024dreamspace} retextures captured real-world indoor scenes from text prompts for immersive VR viewing, while Roomify~\cite{wang2026roomify} further applies spatially grounded style transformations to existing furniture in situ, yet both of them do not generate new layouts. Welti et al.~\cite{welti2026spatial} extend MR with a scale-registered VR replica for inspecting otherwise unreachable viewpoints in interior design. However, these systems primarily support scene modification and general-purpose editing rather than in situ room planning. 

Across both settings, prior immersive authoring work supports direct interaction or AI-assisted refinement. However, no existing system combines iterative furniture planning with a controllable blend of real and virtual views, allowing users to evaluate layout proposals against the actual room throughout layout generation, inspection, and revision.

\begin{figure*}[t]
    \centering
    \includegraphics[width=\textwidth]{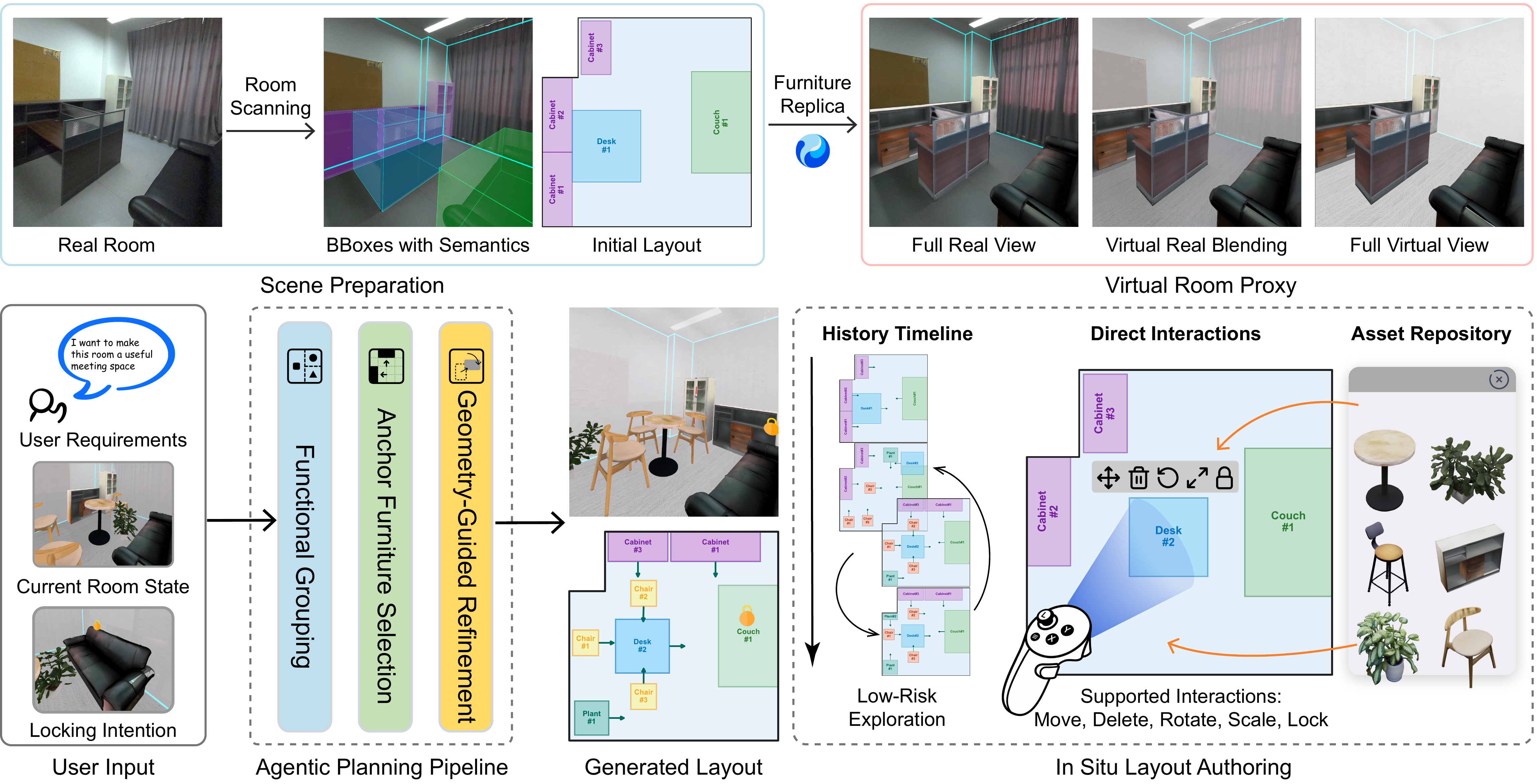}
    \caption{System overview of \SysName{}. The user scans the physical room to construct a spatial representation, then invokes the skill-guided layout agent to generate candidate layouts. Results are presented through a virtual room proxy for in situ observation, comparison, and iterative editing via direct manipulation or language-guided instructions with intent locking.}
    \label{fig:method_system}
\end{figure*}

\subsection{Automated Indoor Layout Synthesis}

Learning-based methods for indoor layout synthesis typically follow a condition-then-generate paradigm~\cite{li2019grains,paschalidou2021atiss,para2023cofs,tang2024diffuscene,leimer2022layoutenhancer,feng2025casagpt}. One-shot generators such as GRAINS~\cite{li2019grains} produce complete scenes from floor plans or object sets, while rearrangement methods such as LEGO-Net~\cite{wei2023lego} and DeBaRA~\cite{maillard2024debara} refine existing configurations through score-based denoising. Notably, LEGO-Net completes the entire rearrangement in a single inference pass without accepting any user input, so users cannot selectively lock approved placements across iterations. DeBaRA addresses this issue with an inpainting mask that preserves specified object attributes during denoising. Yet its conditioning interface is limited to object-semantic categories, providing no language channel for expressing spatial preferences. 
Several generation methods also incorporate room boundary geometry: Forest2Seq~\cite{sun2024forest2seq} encodes floor plans using binary masks, and GLTScene~\cite{li2024gltscene} refines object alignment using wall-relative local coordinates atop a mask-conditioned global placement, thereby enabling explicit handling of polygonal boundaries through geometric structure. However, none of them accept users' language input during the generation process, and existing training datasets~\cite{fu20213d,zhong2025internscenes} are overwhelmingly composed of rectangular rooms, which limits the generalizability of these data-driven methods to the diverse non-rectangular room boundaries commonly found in real-world homes.

LLM- and VLM-based systems extend layout generation with language-guided semantic control. Holodeck~\cite{yang2024holodeck} and related open-vocabulary methods~\cite{gumin2025procedural,sun2025hierarchically} use language to specify scene content before solving for a layout. LayoutVLM~\cite{sun2025layoutvlm} couples vision-language reasoning with differentiable optimization to enforce physical plausibility, and SceneWeaver~\cite{yang2025sceneweaver} frames synthesis as an agentic refinement process with iterative self-evaluation. These approaches substantially improve semantic controllability, yet they share two limitations. First, users can only intervene through text prompts on a 2D screen, which makes it difficult to convey spatial intentions rooted in first-person perception, such as making a room feel less crowded or preserving sightlines. Second, room geometry is commonly parameterized as a fixed width$\times$height rectangle in most LLM- or VLM-based methods, leaving no natural way to describe polygonal boundaries with a variable number of segments.

Across both families of methods, layouts are evaluated only by users after generation is complete, and the interaction channels remain non-immersive. Our system instead places layout generation within a physically grounded human-AI planning loop, in which users evaluate and revise proposals in the target room, preserve approved placements across iterations, and plan directly over irregular polygonal room boundaries.

\subsection{General-Purpose Agents for Domain-Specific Tasks}

Recent work has adapted general-purpose agents such as Codex and Claude Code to specialized domains. Instead of producing a single answer, these agents work within domain-specific environments and revise their outputs based on execution results. ParaCodex~\cite{kaplan2026paracodex} applies this approach to OpenMP GPU offloading, where correctness checks and profiling results guide Codex as it improves parallel programs. LLMoxie~\cite{setiawan2026llmoxie} extends Claude Code with a Plugin--Agent--Skill structure designed around scientific software development. Together, these systems show that general-purpose agents can be adapted through domain guidance and execution feedback.

Our system applies this pattern to real-room layout planning through a room-layout design skill, which encodes the design principles we formulate and the spatial knowledge needed to interpret the target room. The skill-guided layout agent generates and refines layout updates for in situ authoring using a large language model and feedback from geometric validation.

\begin{figure*}[t]
    \centering
    \includegraphics[width=\textwidth]{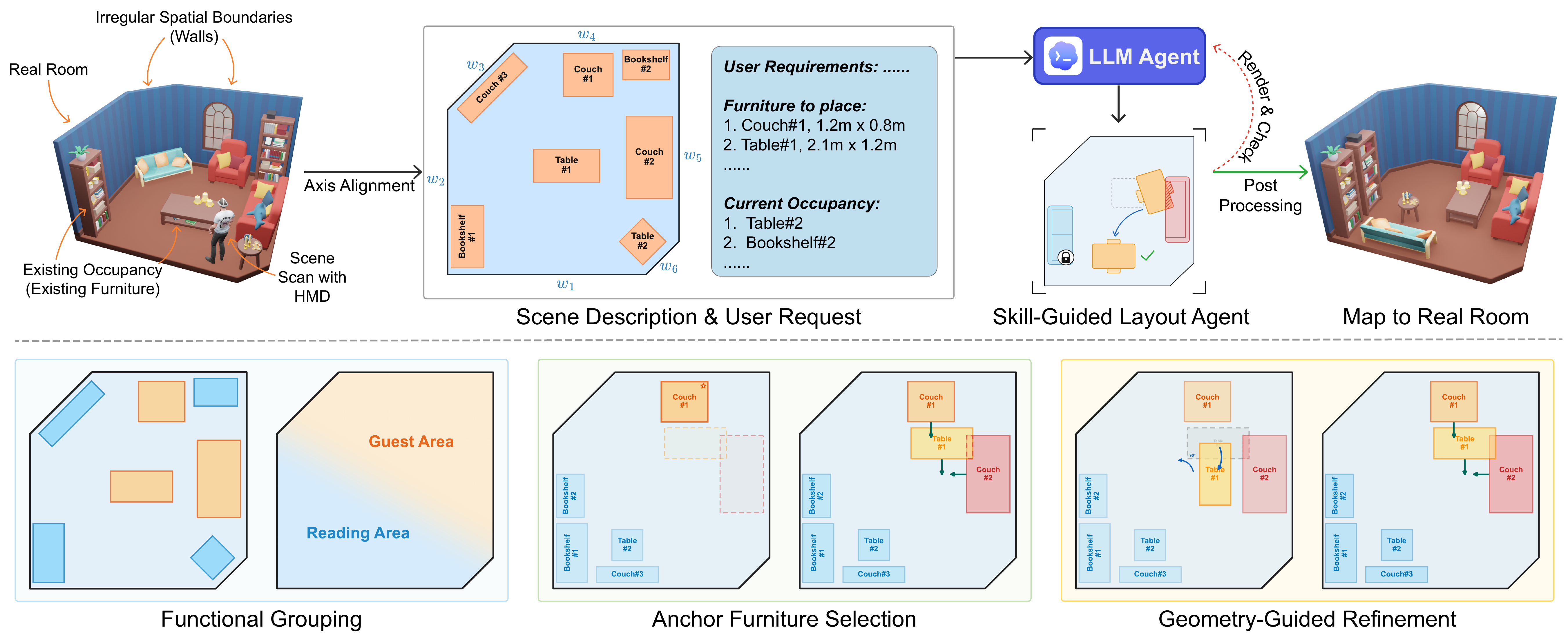}
    \caption{Skill-guided layout generation in \SysName{}. \textbf{Top row:} an HMD scan captures the room's irregular boundaries and current furniture occupancy. After axis alignment, these scene data are combined with the user requirements and furniture to be placed and supplied to an LLM agent. Guided by the room-layout design skill, the agent generates complete candidate layouts and iteratively renders, checks, and revises them while preserving locked furniture poses; the selected layout is then post-processed and mapped back to the real room. \textbf{Bottom row:} three complementary design principles: functional grouping establishes activity regions and related furniture groups; anchor furniture selection places a function-defining item before arranging associated furniture; and geometry-guided refinement adjusts orientations and positions to improve spatial relationships and resolve geometric violations.}
    \label{fig:method_algorithm}
\end{figure*}

%% file: text/3_method.tex
\section{\SysName{}}

\SysName{} is an in situ authoring system for real-room layout design organized around a shared layout state. The system maintains the user's evolving spatial proposal as a layout state that records the furniture set, furniture dimensions and poses, and locked placements. The virtual room proxy, direct manipulation, language-guided editing, and the skill-guided layout agent all operate on this state and therefore act on the same layout proposal.

The user first loads the target room scanned by a head-mounted display. \SysName{} then presents the layout state at physical scale through a virtual room proxy registered to the room. The user can adjust individual furniture items with direct manipulation, express higher-level changes in natural language, and preserve accepted placements through intent locking. The skill-guided layout agent combines the layout state, user requirements, and room geometry to generate a complete layout update, which is returned to the virtual room proxy for further in situ evaluation and editing.

\subsection{In Situ Authoring Interface}
\label{sec:authoring}

To support iterative layout planning grounded in the user's physical room, we design an in situ authoring interface that unifies three complementary interaction modalities: a virtual room proxy for observing and comparing candidate layouts against the real environment, direct manipulation with visual feedback for local spatial adjustments, and language-guided editing with intent locking for expressing high-level planning intents while preserving previously confirmed decisions. These three components form an authoring loop in which users generate, evaluate and refine layouts without leaving the physical room. All three mechanisms read from and update the layout state: the virtual room proxy presents the active state in the target room, direct manipulation changes it immediately, and language-guided editing requests a new state from the layout agent.

\subsubsection{Virtual Room Proxy}
\label{sec:virtual_room_proxy}

A central challenge in real-room planning is that users must judge whether a candidate layout suits the actual dimensions, lighting conditions, doorway positions, and existing object distribution of their space. Abstract room boundaries or fully virtual previews require users to mentally map proposals onto the real room, increasing cognitive load and reducing evaluation accuracy. To alleviate this mapping burden, we design the virtual room proxy, a spatially registered virtual layout carrier that is aligned with the real room's walls, floor, and architectural features. Rather than presenting an isolated virtual environment, the proxy projects candidate furniture placements into the physical room context, allowing users to perceive what a proposed layout would look like if realized in their actual space. To further enhance visual fidelity, the system employs an image-to-mesh generation model to reconstruct 3D meshes of existing furniture from captured images. These reconstructed replicas replace abstract bounding-box placeholders with faithful representations of the user's actual belongings, so that repositioned items and reused assets appear realistic within the proxy.

The virtual room proxy supports three complementary visual states that vary how much of the physical environment remains visible while keeping the candidate furniture layout spatially registered:
\begin{itemize}[leftmargin=10pt, nosep]
    \item \textbf{Full Real View.} The live passthrough remains fully visible, while candidate virtual furniture items are overlaid at their registered positions in the physical room, allowing users to assess the proposed layout directly against the actual surroundings.
    \item \textbf{Virtual Real Blending.} The spatially registered virtual room proxy is composited with the passthrough at an adjustable opacity, reducing visual interference from the physical room while preserving enough real-world context to compare the proposed and current layouts.
    \item \textbf{Full Virtual View.} The passthrough feed is fully replaced by the virtual room proxy, immersing users in a fully virtual representation of the candidate layout while preserving the physical room's scale and spatial registration.
\end{itemize}
Users can switch among these states and adjust the blending level, creating a smooth perceptual transition from an MR overlay grounded in the physical room to a fully virtual layout preview.

Beyond visualization, the virtual room proxy serves as the spatially registered display and interaction surface for the layout state. Direct manipulations update this shared state through the proxy, while layouts produced by language-guided editing are written back to the same state and re-rendered in place.

\subsubsection{Direct Manipulation and Visual Feedback}
\label{sec:direct_manipulation}

Although the skill-guided layout agent can produce high-level layout proposals, room planning frequently involves fine-grained adjustments that are difficult to express verbally. To support such local refinements, the authoring interface provides direct manipulation within the virtual room proxy. Users select furniture via controller ray casting and can then perform four operations: single-hand dragging for floor-plane translation, two-handed gestures for vertical-axis rotation with discrete angular snapping, deletion to remove unwanted items, and a locking action that marks the item as confirmed so that it is preserved during subsequent re-planning. In addition to editing agent-generated furniture, users can browse an asset panel to add new furniture items to the scene; user-added assets further support uniform scaling, giving users full control over object size. Real-time visual feedback, including selection highlights, operation-mode indicators, boundary warnings, and lock icons, keeps users informed of their interaction state throughout the process.

The system also maintains a history of snapshots of the complete layout state, allowing users to switch between agent-generated results and manually edited states, and to undo changes as needed. Restoring a snapshot jointly restores its furniture set, object dimensions, continuous poses, and lock status. This ensures that exploration remains low-risk: users can freely experiment knowing that any prior state is recoverable.

\subsubsection{Language-Guided Editing and Intent Locking}
\label{sec:language_editing}

Direct manipulation excels at local, spatially precise edits but is poorly suited for expressing higher-level semantic goals, such as ``keep the bed away from the door,'' or ``preserve this table's position and re-plan everything else.'' To handle such intents, \SysName{} incorporates a language-guided editing channel that allows users to express functional and spatial preferences in natural language, complementing the fine-grained control of direct manipulation. Users can provide modification intents through either a text input field or speech recognition. The system converts the input into natural-language requirement description that is forwarded to the skill-guided layout agent together with the current layout state.

When users are satisfied with specific furniture placements obtained through either direct manipulation or a previous generation round, they can lock those items in place. The system stores lock status with each furniture item in the layout state. For every generation request, the layout agent receives all furniture identities, dimensions, continuous poses, and lock status through the visual and structured state inputs. Locked poses act as hard constraints that the updated layout must preserve, whereas any unlocked furniture remains eligible for rearrangement. With intent-locking mechanism, multi-round refinement remains cumulative because each iteration builds on approved decisions.

\subsection{Spatial Scene Representation}
\label{sec:scene_representation}

\SysName{} takes the scene understanding data obtained from the head-mounted display (e.g., Meta Quest~3), which provides two categories of spatial information: spatial boundaries, comprising walls, floor, ceiling, and door frames that define the room's geometric envelope, and entity occupancy, comprising semantic labels and 3D bounding boxes of furniture already present in the room. \SysName{} encodes the room boundary as an ordered collection of individually numbered wall segments, allowing the same representation to describe rectangular, L-shaped, and other polygonal rooms with varying numbers of walls.

At authoring step $t$, \SysName{} represents the evolving furniture arrangement as the layout state $\mathcal{S}_t=\{(f_i,\mathbf{p}_i,\ell_i)\}_{i=1}^{N_t}$, where $f_i$ records a furniture item's identity, category, dimensions, and visual asset, $\mathbf{p}_i=(x_i,z_i,\theta_i)$ is its continuous footprint-center position and front-facing yaw, and $\ell_i\in\{0,1\}$ records whether its pose is locked. This shared representation carries the furniture geometry, arrangement, and preservation decisions used by both the authoring interface and the layout agent.

A room captured in the HMD coordinate system typically has an arbitrary orientation that does not align with the coordinate axes.
To provide the layout agent with a consistent spatial reference for reasoning, \SysName{} constructs an axis-normalized view by rotating the spatial representation as a whole around the centroid of the room boundary until its longest wall segment is parallel to the nearest X or Z axis.
Formally, let $\alpha$ be the direction of the longest wall in the original X/Z plane. Its angular deviation from the nearest axis and the corresponding rotation are
\begin{equation*}
    \delta=((\alpha+45^\circ)\bmod 90^\circ)-45^\circ, \qquad \phi=-\delta.
\end{equation*}
For the centroid $\mathbf{c}$ of the room boundary, any point $\mathbf{p}^{w}$ in the original coordinate frame, and the planar rotation matrix $R(\phi)$, the normalized position and furniture orientation are
\begin{equation*}
    \mathbf{p}^{v}=\mathbf{c}+R(\phi)(\mathbf{p}^{w}-\mathbf{c}), \qquad \theta^{v}=\theta^{w}-\phi.
\end{equation*}

The same rotation is applied to the boundary, walls, openings, and furniture poses, giving the layout agent aligned visual and structured evidence for interpreting their spatial relationships. The representation preserves the room's metric scale and complete polygon geometry, including concave boundaries and oblique wall segments. Positions are expressed in meters at furniture-footprint centers, with $0^\circ$ facing $+Z$, $90^\circ$ facing $+X$, and positive yaw rotating clockwise in the supplied top-down view.

\subsection{Skill-Guided Layout Agent}
\label{sec:llm_planning}

To sustain the in situ authoring process, \SysName{} uses a general-purpose LLM agent guided by the room-layout design skill as its layout back end. During initial generation and later language-guided edits, the layout agent uses the existing user requirements and layout state together with scanned room information to produce a complete layout update. Furniture identities, dimensions, and poses provide the spatial context, while locked poses encode user-confirmed decisions as hard constraints.

The layout agent receives paired visual and structured inputs. The visual input presents an axis-normalized top-down view of the room boundary, walls, openings, furniture footprints, and front directions, and distinguishes locked from unlocked furniture. The structured input describes the same scene with precise boundaries, openings, furniture identities, dimensions, continuous poses, and lock status. Both inputs encode the layout state: visual evidence supports reasoning about room shape, regional relations, and overall composition, while structured evidence provides the exact geometry needed to predict furniture poses in continuous coordinates. The agent also receives the user requirements and expresses each layout update through furniture footprint centers and front directions in the X/Z coordinates defined by the spatial scene representation.

\subsubsection{Room-layout Design Skill}

The room-layout design skill adapts a general-purpose LLM agent to real-room layout design. Drawing on established interior-design guidance~\cite{kilmer2014designing,panero1979human}, we formulate three complementary principles for agent-based real-room layout generation. These principles translate established knowledge about functional relations, spatial organization, circulation, and human dimensions into executable guidance for reading spatial evidence, expressing a complete layout in continuous coordinates, and revising it through rendering and geometric checks. The skill operationalizes this formulation through functional grouping, anchor furniture selection, and geometry-guided refinement.

\textbf{Functional grouping.} Functional grouping organizes spatial relations at both the region and furniture scales. The agent establishes coherent activity regions from the furniture set, room shape, and user requirements, then arranges related furniture within each region according to functional adjacency and orientation. A bed and nightstands form a sleeping group, seating can be arranged around a table, and a sofa, coffee table, and media furniture can form a clear use relation. Functional grouping, therefore, coordinates room-level region allocation with local furniture relations.

\textbf{Anchor furniture selection.} Anchor furniture selection identifies the large or function-defining item in each region. The agent places this item first according to the room boundary, openings, and functional orientation, then arranges related furniture around it. This principle develops the layout from its main spatial relations while allowing furniture-to-wall relations to follow functional grouping and access needs.

\textbf{Geometry-guided refinement.} Geometry-guided refinement uses the room polygon, openings, furniture poses, locked poses, rendering, and geometric checks to revise the layout. The agent gives priority to entrances, main routes, furniture fronts, and access between functional regions while also checking orientations, furniture pairings, and overall composition. It adjusts the relevant poses or reorganizes unlocked furniture according to the spatial scope of a problem while preserving locked poses.

\subsubsection{Adaptive Layout Generation and Refinement}

The room-layout design skill provides spatial evidence, design principles, a layout representation based on continuous coordinates, and rendering and geometric checks, but does not impose a fixed execution sequence. For each initial generation or language edit, the agent decides how to order scene interpretation, functional planning, complete layout generation, checking, revision, and candidate selection, and can revisit earlier spatial decisions based on intermediate results. The agent checks the geometry of each candidate layout and ensures that all locked poses are preserved.

A common execution path proceeds from coarse to fine. The agent identifies functional regions and main routes from the room shape, openings, existing layout, and user requirements, selects anchor furniture for each region, and arranges related furniture around these anchors to form a complete candidate layout. The skill treats this path as guidance rather than a prescribed series of stages. During a language edit, the agent can instead begin from the existing layout and change the affected spatial relations; after checking a candidate, it can return to functional grouping, anchor placement, or the overall composition.

For a complete candidate layout, the agent can use rendering and geometric checks as needed to assess the furniture inventory, boundary compliance, overlaps, circulation, functional relations, and overall composition. Based on the results, the agent can retain the candidate, adjust individual items, reorganize a functional region, replace the overall arrangement, generate another candidate, or end the process. It can continue revising a candidate after the geometric conditions are satisfied to improve function and composition, or directly select the first candidate when it satisfies both the geometric conditions and the design principles. The selected valid layout then updates the layout state and is presented through the virtual room proxy.

%% file: text/4_implementation.tex
\section{Implementation Details}

We implemented \SysName{} using the Meta Quest 3, which offers built-in SLAM tracking and a depth sensor to support MR experiences. The \SysName{} interface was developed on a local PC (Intel Core i9-14900K CPU, 128GB RAM) using Unity 6000.1.10f1.

For environment capture, the system relies on the built-in scene understanding module~\cite{meta2026xrallinone} of Meta Quest~3, requiring no external scanning devices. The captured scene is converted into the spatial representation described in Section~\ref{sec:scene_representation}. For each detected furniture item, \SysName{} uses Hunyuan3D Studio~\cite{lei2025hunyuan3dstudioendtoendai} to reconstruct a textured 3D model from a single image.

Each layout generation is executed with Codex~0.147.0 in a temporary Docker environment with file-system access limited to the inputs for the current task, preventing information leakage from unrelated files on the host system. The agent uses GPT-5.6 Sol~\cite{openai2026gpt56sol} with high reasoning effort as foundation model. The agent applies the room-layout design skill to the user request in the context of the current scene representation. The agent predicts the position and orientation of each movable furniture item in continuous coordinates, while treating locked furniture as fixed geometry.

During generation, each candidate layout undergoes deterministic validation for completeness and geometric validity. Codex reviews these messages to check whether the layout is practical and coherent. Once selected, the layout is written back to the shared layout state and displayed in the virtual room proxy.

%% file: text/5_technical_eval.tex
\section{Technical Evaluation}

We evaluate \SysName{}'s layout generation pipeline through quantitative comparison with existing automated layout synthesis methods, qualitative visual analysis, and a human perceptual evaluation study.
The goal is to assess how well skill-guided agentic planning produces physically plausible and semantically coherent furniture arrangements for rooms with non-rectangular boundaries.
The full-system user experience, including the MR authoring interface and iterative refinement workflow, is evaluated in the system study (\S\ref{sec:user_study}).

\subsection{Quantitative Evaluation}

\subsubsection{Baselines}
We compare against two representative methods from distinct methodological families.

\textbf{LayoutVLM}~\cite{sun2025layoutvlm} is a VLM-based open-universe layout generation method that predicts numerical object poses and spatial relations from visually marked scene renderings, then refines placements through differentiable optimization that jointly minimizes physics-based collision objectives and semantic relational objectives.
It represents the current state of the art in language-guided layout generation.
Because the original method parameterizes rooms with four axis-aligned walls, \emph{LayoutVLM-Original} uses the minimum bounding rectangle of each non-rectangular test room as its room input. We additionally use a polygonal-boundary adaptation, denoted \emph{LayoutVLM-PolyRepair}, that supplies the true room boundary and adds a full-footprint polygon constraint to the joint placement optimization. Implementation details of \emph{LayoutVLM-PolyRepair} are provided in the supplementary material.

\textbf{GLTScene}~\cite{li2024gltscene} is a data-driven method that employs a global-to-local transformer architecture trained on the 3D-FRONT dataset~\cite{fu20213d}.
A global placement transformer predicts coarse furniture positions conditioned on a floor-plan mask, while a local alignment transformer refines orientations relative to the nearest wall.
GLTScene is specifically designed for general polygonal room boundaries, making it the most relevant learning-based baseline for evaluation.

\subsubsection{Evaluation Metrics}
We adopt six metrics from the evaluation protocols of LayoutVLM~\cite{sun2025layoutvlm} and GLTScene~\cite{li2024gltscene} to evaluate physical plausibility, semantic coherence, and wall-alignment quality. Specifically, CF and IB measure collision avoidance and boundary compliance, PC and RC assess whether semantically related furniture items are positioned and oriented in functionally coherent ways based on LLM judgments from top-down and side-view renderings, PSA combines semantic quality with penalties for physical infeasibility, and AE measures each item's orientation consistency with respect to its nearest wall. Detailed metric definitions and the exact AE formulation are provided in the supplementary materials.

\subsubsection{Experimental Setup}
We evaluate on 43 rooms sampled from the 3D-FRONT dataset~\cite{fu20213d}, nearly all of which have non-rectangular room boundaries (e.g., L-shaped or otherwise polygonal boundaries).
For each room, the furniture set comprises all original items in the scene excluding lighting fixtures, preserving the ground-truth object categories and dimensions.
Each method generates a single layout per room.

In this evaluation, \SysName{} and LayoutVLM-PolyRepair both use GPT-5.6 Sol~\cite{openai2026gpt56sol} with high reasoning effort. \SysName{} receives the complete room-layout design skill and a rendering of the current layout. LayoutVLM-Original retains the original manuscript configuration with minimum-bounding-rectangle room input.
For GLTScene, we follow the original evaluation protocol: the model receives the polygonal floor-plan mask directly as its room boundary input, and we adopt a conditional generation setting in which ground-truth furniture categories and sizes are provided, so that the model predicts only placement positions and orientations.
PC, RC, and the layout-criteria-match ratings used by PSA are computed using the prompt templates of LayoutVLM~\cite{sun2025layoutvlm}, with Gemini~3~Flash~\cite{google2025gemini3flash} rating each rendered layout.

\subsubsection{Results}

\begin{table}[t]
\centering
\small
\caption{Quantitative comparison of layout generation quality on non-rectangular rooms. PC, RC, and the layout-criteria-match ratings used by PSA are evaluated by Gemini~3~Flash. Best results are in \textbf{bold}; second-best are \underline{underlined}.}
\label{tab:quantitative}
\setlength{\tabcolsep}{3pt}
\begin{tabular}{l cccccc}
\toprule
Method & CF~$\uparrow$ & IB~$\uparrow$ & PC~$\uparrow$ & RC~$\uparrow$ & PSA~$\uparrow$ & AE~$\downarrow$ \\
\midrule
GLTScene~\cite{li2024gltscene}       & 0.8075   & \underline{0.9532}   & 54.72   & 50.28   & \underline{39.53}   & 0.0783   \\
LayoutVLM-Original~\cite{sun2025layoutvlm}    & 0.5338   & 0.5529   & 45.72   & 38.77   & 13.29   & 0.3291   \\
LayoutVLM-PolyRepair~\cite{sun2025layoutvlm}  & \underline{0.8548}   & 0.7295   & \underline{62.56}   & \underline{54.72}   & 36.30   & \underline{0.0740}   \\
\SysName{}                            & \textbf{0.9731}   & \textbf{1.0000}   & \textbf{65.60}   & \textbf{58.12}   & \textbf{55.78}   & \textbf{0.0083}   \\
\bottomrule
\end{tabular}
\end{table}

Table~\ref{tab:quantitative} reports the aggregate results on the 43-room benchmark. With true-polygon input and full-footprint boundary repair, LayoutVLM-PolyRepair improves all six metrics over its original version and provides the main LayoutVLM comparison for irregular rooms. The two polygon-aware baselines show different strengths: GLTScene has the highest baseline boundary compliance (IB~0.9532), whereas LayoutVLM-PolyRepair has the highest baseline PC and RC but a lower IB of 0.7295. Compared with LayoutVLM-PolyRepair, \SysName{} increases CF from 0.8548 to 0.9731, improves PC and RC by 3.04 and 3.40 points, and reduces AE from 0.0740 to 0.0083. It also reaches complete boundary compliance, exceeding GLTScene on its strongest metric.

The combined improvement is most clearly reflected in PSA. \SysName{} reaches 55.78, compared with the strongest baseline result of 39.53. Because PSA combines semantic quality with collision and boundary compliance, this result shows that \SysName{} produces coherent furniture arrangements while satisfying the physical constraints of polygonal rooms. Additional per-room-type results are provided in the supplementary materials.

\subsection{Perceptual Evaluation}

To complement the automatic metrics, which primarily capture geometric correctness, we conducted a perceptual evaluation study in which independent human raters assessed layout quality that automated measures cannot fully capture, such as functional coherence and holistic preference.

\subsubsection{Study Design}
We selected 9 representative non-rectangular rooms from the evaluation set, covering a range of floor-plan complexities (e.g., L-shaped) and room areas.
GLTScene, LayoutVLM-PolyRepair, and \SysName{} each produced one layout per room, yielding $9 \times 3 = 27$ rooms in total.
Every room was presented as a pair of rendered images: a \emph{top-down view} showing the overall spatial arrangement and a \emph{perspective view} conveying the layout from a human viewpoint.

Participants rated each layout on three dimensions using a 7-point Likert scale ($1$ = very poor, $7$ = excellent):
\begin{itemize}[leftmargin=10pt,nosep]
    \item \textbf{Physical Plausibility}: whether all furniture lies within the room boundary, free of collisions and unreasonable floating.
    \item \textbf{Functional Coherence}: whether furniture groupings form logical clusters and functional zones are clearly delineated.
    \item \textbf{Overall Preference}: holistic satisfaction integrating physical plausibility, functional coherence, and overall design quality.
\end{itemize}

The study followed a within-subjects design: every participant evaluated all 27 layouts.
Method labels were anonymized, with the label-to-method mapping randomized per participant.

\begin{figure}[t]
    \centering
    \includegraphics[width=0.8\columnwidth]{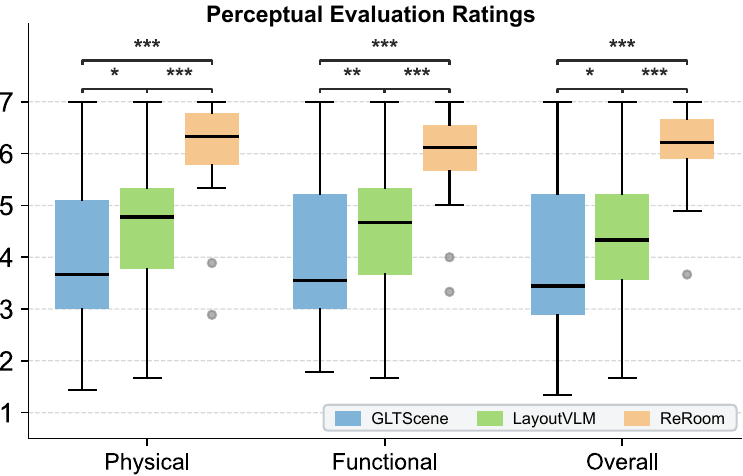}
    \caption{Perceptual evaluation ratings ($N{=}45$; 7-point Likert scale) for \textcolor{PerceptGLTScene}{\textbf{GLTScene}}, \textcolor{PerceptLayoutVLM}{\textbf{LayoutVLM-PolyRepair}}, and \textcolor{PerceptReRoom}{\SysName{}}.}
    \label{fig:perceptual_evaluation}
\end{figure}

\subsubsection{Participants and Procedure}
We recruited 45 participants (32 male, 71.1\%; 13 female, 28.9\%; aged 21--65, $M=34.0$, $SD=12.6$, $\mathit{Mdn}=27$).
After providing informed consent and basic demographic information, each participant reviewed a brief tutorial containing one annotated example layout as a calibration anchor (excluded from analysis).
Participants then rated all 27 layouts on the 3 dimensions described above.

\subsubsection{Results}
We analyzed each dimension separately using two-sided pairwise Wilcoxon signed-rank tests on each participant's mean rating across the nine rooms. Figure~\ref{fig:perceptual_evaluation} summarizes the ratings. The perceptual advantage of \SysName{} was both multidimensional and robust. Thirty-four of 45 participants assigned it higher mean ratings than both baselines on all three dimensions, and \SysName{} achieved the highest mean overall-preference rating in eight of the nine rooms while tying for highest in the remaining room. LayoutVLM-PolyRepair exceeded GLTScene in physical plausibility ($p=.015$), functional coherence ($p=.005$), and overall preference ($p=.018$), whereas \SysName{} exceeded both baselines across all three dimensions (all $p<.001$). These patterns show that \SysName{}'s overall preference advantage reflects concurrent gains in physical plausibility and functional coherence that persist across participants and room instances.

%% file: text/6_user_study.tex
\begin{figure*}[t]
    \centering
    \includegraphics[width=\textwidth]{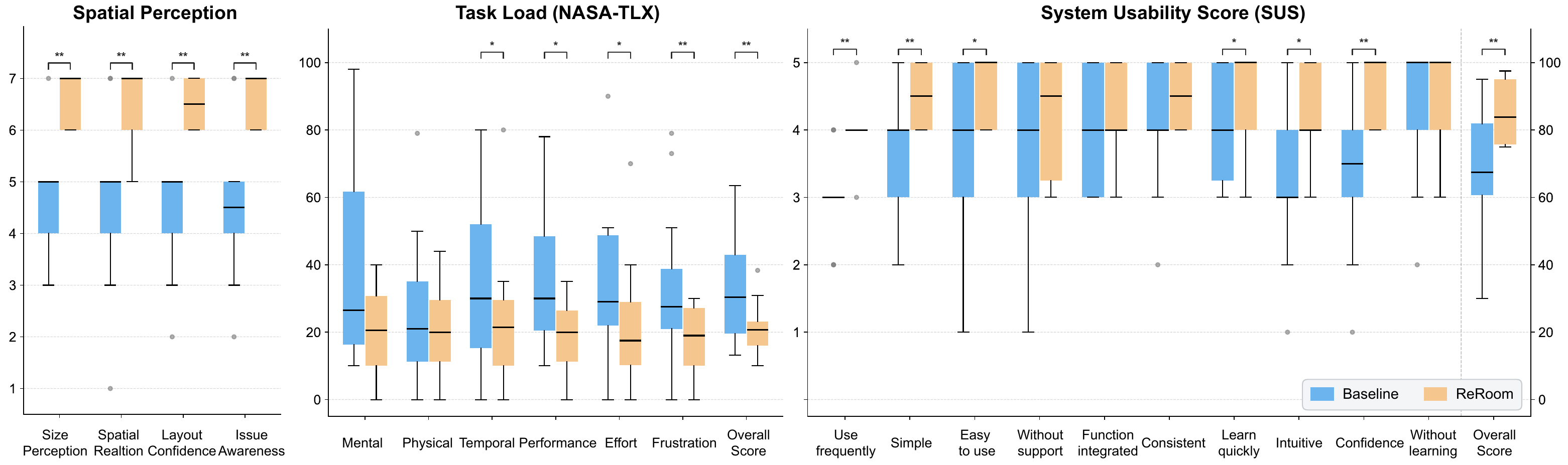}
    \caption{Questionnaire results ($N=14$) from the system usability study comparing the \textcolor{UserStudyBaseline}{\BaselineName{}} (VR, off-site) and \textcolor{UserStudyReRoom}{\SysName{}} (MR, in situ). \textbf{Left:} Four custom items on spatial perception and in situ evaluation quality (7-point Likert; higher is better). \textbf{Center:} NASA-TLX workload subscales (0 to 100; lower is better). \textbf{Right:} SUS composite score (1 to 5 for each item, 0 to 100 overall; higher is better). Whiskers extend to $1.5\times$ the interquartile range; gray dots denote outliers. Significance levels: {*}\,$p<.05$, {**}\,$p<.01$, {***}\,$p<.001$.}
    \label{fig:system_usability_study}
    \end{figure*}

\section{System Usability Study}
\label{sec:user_study}

While our technical evaluation confirms that \SysName{} generates physically plausible and functionally coherent layouts, effective interior planning ultimately depends on how well users can evaluate and refine these suggestions within their target environment.

Prior work~\cite{zhang2024vrcopilot,hou2025echoladder} has shown the value of AI assistance in immersive design; building on this, our user study investigates whether keeping generated proposals spatially registered to the physical room (in situ MR) improves spatial perception and system usability. To isolate the benefits of in situ grounding, we conducted a within-subjects user study comparing \SysName{} against \BaselineName{}, a VR-only baseline that shares identical generation and interaction capabilities but lacks physical room context.

\begin{figure}[t]
    \centering
    \includegraphics[width=\columnwidth]{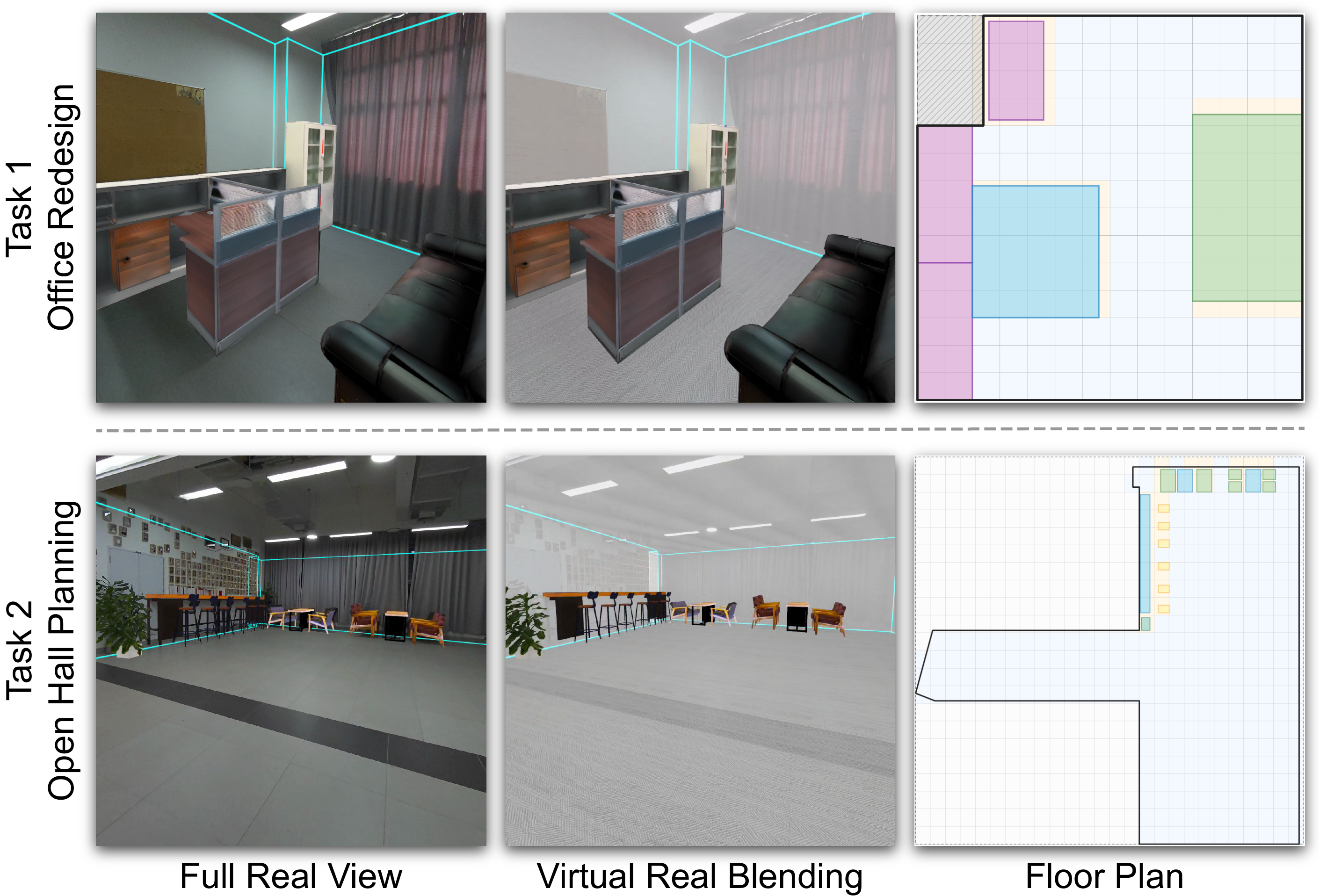}
    \caption{Task environments used in the system usability study. \textbf{Top row:} Task~1 (office redesign). \textbf{Bottom row:} Task~2 (open hall planning). Each row shows, from left to right, the \emph{Full Real View}, in which candidate virtual furniture is overlaid on the physical-room passthrough; \emph{Virtual Real Blending}, in which the spatially registered virtual room proxy is composited with the passthrough; and the \emph{Floor Plan} representation of the task initial layout state.}
    \label{fig:exp_task_view}
\end{figure}

\subsection{Study Design}

\subsubsection{Conditions and Participants}
We compared two conditions in a within-subjects design:

\textbf{\SysName{} (MR, in situ).} The full system described in \S\ref{sec:authoring}, where participants plan and refine furniture layouts while physically standing in the target room, with the virtual room proxy spatially registered to the real environment.

\textbf{\BaselineName{} (VR, off-site).} A system that provides the same skill-guided layout generation backend (\S\ref{sec:llm_planning}) and the same interaction modalities with direct manipulation, language-guided editing, and intent locking. \BaselineName{} presents the room and furniture in a purely virtual environment without physical room context. To ensure participants had enough space for physical movement, the \BaselineName{} sessions were conducted in a sufficiently large open space. We also provided joystick-based smooth locomotion as an additional navigation option, allowing participants to traverse the virtual scene without requiring physical walking.

We recruited 14 participants (12 male, 2 female; aged 18--27, $M=23.6$, $SD=2.5$); 9 had prior experience with AR/MR/VR systems. None had professional interior design training.

\subsubsection{Tasks and Procedure}
Each participant completed two room-planning tasks with each system, yielding 4 tasks in total. Figure~\ref{fig:exp_task_view} shows the task environments.

\textbf{Task~1:} In a near-rectangular room, participants redesigned a double-occupancy office into either a single-person office or a discussion space. This task tested planning under tight spatial constraints with existing furniture to consider.

\textbf{Task~2:} In a larger, non-rectangular hall, participants designed a new activity space from scratch. This task tested open-ended planning in an irregular room geometry.

To mitigate order effects, the system presentation order was counterbalanced: half of the participants began with \SysName{} and the other half with \BaselineName{}.
Each participant received a brief tutorial on the assigned system before starting the tasks.
After completing all four tasks, participants completed a post-study questionnaire and participated in a semi-structured interview.

\subsubsection{Measures}
The post-study questionnaire comprised three components.
(1)~The \emph{NASA Task Load Index} (NASA-TLX)~\cite{hart1988development} assessed perceived workload across six subscales: Mental Demand, Physical Demand, Temporal Demand, Effort, Performance, and Frustration, each rated on a 0 to 100 scale.
(2)~The \emph{System Usability Scale} (SUS)~\cite{brooke1996sus}, a ten-item instrument scored on a 5-point Likert scale, yielded a composite usability score ranging from 0 to 100.
(3)~Six custom 7-point Likert-scale items targeted spatial perception and in situ evaluation quality.
Four items were shared between conditions:
\emph{Size Perception} (``I could accurately assess the real physical size of virtual furniture''),
\emph{Spatial Relation} (``I could clearly judge spatial relationships between virtual furniture and existing room structures such as doorways''),
\emph{Layout Confidence} (``After completing the layout, I am confident the design could be faithfully reproduced in the real room''),
and \emph{Issue Awareness} (``The system allowed me to intuitively discover layout problems, such as blocked circulation paths, and make corrections'').
Two additional items were administered only for \SysName{}:
\emph{Visual Comparison} (``The visual state switching feature greatly helped me evaluate virtual layouts against the real room'')
and \emph{Environmental Awareness} (``The visual state switching feature let me focus on the virtual layout while maintaining awareness of the physical world and feeling safe'').

All comparisons between \SysName{} and \BaselineName{} used two-sided paired Wilcoxon signed-rank tests. We report matched-pairs rank-biserial correlation ($r_{\mathrm{rb}}$) as the effect size.

\subsection{Quantitative Findings}

The questionnaire results in Figure~\ref{fig:system_usability_study} reveal a coherent pattern. \SysName{} strengthened participants' connection between virtual proposals and physical constraints, and this advantage extended to the overall experience of carrying out the planning task.

\emph{Spatial Perception.} The clearest difference emerged across all four shared measures, with consistently large paired effects (all $p\leq.0022$, $r_{\mathrm{rb}}\geq.95$). This pattern spans two stages of planning: forming spatial judgments about a proposal and deciding whether it would work in the real room. In situ grounding therefore helped participants translate visual inspection into layout decisions.

\emph{Task Load (NASA-TLX).} \SysName{} produced a lower composite Raw TLX score than \BaselineName{} ($\mathit{Mdn}=20.75$ vs.\ $30.42$; $p=.0017$, $r_{\mathrm{rb}}=.89$). The subscale pattern shows that this reduction centered on how participants carried out and appraised the planning task. This concentration shows that in situ grounding made layout evaluation more manageable.

\emph{System Usability Score (SUS).} \SysName{} also achieved a higher composite SUS score ($\mathit{Mdn}=83.75$ vs.\ $67.50$; $p=.0022$, $r_{\mathrm{rb}}=1.00$). Across the SUS questions, the advantage covers both lower interaction friction and greater confidence and willingness to use the system. This pattern shows that the benefit of in situ grounding extended from spatial evaluation to the overall planning experience.

\emph{Visual State Switching (\SysName{} only).} Participants consistently associated visual state switching with closer comparison between virtual layouts and the real room (\emph{Visual Comparison}: $M=6.64$, $SD=0.50$) while maintaining awareness of their physical surroundings (\emph{Environmental Awareness}: $M=6.43$, $SD=0.65$). This pairing captures the intended balance between detailed inspection and physical-environment awareness.

\subsection{Qualitative Findings}

We conducted semi-structured interviews with all 14 participants after they completed both conditions.
Interviews were recorded, transcribed, and analyzed using thematic analysis.
Two researchers independently coded the transcripts, iteratively discussed emergent codes, and consolidated them into the following themes.
We report representative quotes below. 
The participants are denoted P1--P14.

\subsubsection{In Situ Spatial Awareness and Physical Grounding}
A recurring advantage of \SysName{} was the sense of physical presence and real-world reference it afforded.
P1 noted that the mixed-reality overlay allowed him to ``\textit{feel as though [he] was truly situated in the layout environment}.'' This physical grounding also affected how participants moved through the layout: six participants (P1, P4, P6, P9, P10, and P12) reported greater confidence navigating the scene when the surrounding physical environment remained visible.
P3 found the MR setting ``\textit{more intuitive and practical}'' than the \BaselineName{} condition, explaining that the lack of alignment between the virtual scene and the real room in \BaselineName{} made it ``\textit{easy to misjudge where furniture would actually end up}.''
P5 highlighted that \SysName{} enabled him to ``\textit{immediately see the result and its concrete placement in the real world},'' yielding layouts that were ``\textit{closest to the actual real-world arrangement}.''
P11 similarly valued the ability to ``\textit{combine the virtual layout with reality for an immersive reference},'' and specifically praised the opacity-adjustment feature for providing ``\textit{a reference to the real space}'' that gave him ``\textit{more confidence in judging whether a deletion was reasonable and in further refining the layout}.''

\subsubsection{Efficient Layout Comparison and Evaluation}
Several participants noted that \SysName{} facilitated rapid comparison between proposed layouts and the existing room.
P2 appreciated the ability to ``\textit{quickly compare and place furniture}'' with ``\textit{greater flexibility and a stronger sense of space}.''
P4 found ``\textit{more convenient}'' to work with \SysName{} because direct comparison was always available.
P5 remarked that the in situ feedback made layouts ``\textit{more precise}'' by closing the gap between virtual proposals and physical reality.
P12 noted the visual state switching feature helped him ``\textit{judge whether furniture removal was reasonable}.''

\subsubsection{Multimodal Controls in Iterative Layout Refinement}
Participants combined generation, language input, direct manipulation, and intent locking as different needs arose rather than following a fixed sequence. Participants began with minimal prompts to clarify preferences (P4, P6, P9, P14), then used language for relational edits and direct manipulation for precise placement (P9, P13), consistent with prior XR authoring findings~\cite{zhang2024vrcopilot,lee2025imaginatear}. Participants locked accepted furniture before regenerating unlocked items (P7, P9--P11), preferring explicit locks over language prompts to keep accepted placements unchanged (P9--P11). Regeneration complemented direct manipulation by handling multi-object rearrangements (P8, P13). Overall, participants approximately accepted 75.2\% of system-proposed placements. Notably, in P13's in situ session, a generation round reorganized 38 items and the user made only one subsequent manual adjustment.

\subsubsection{Challenges of Virtual--Physical Coexistence}
Despite the overall positive reception, participants identified a unique challenge arising from the high visual fidelity of the mixed-reality overlay.
P1 reported that virtual furniture could appear ``\textit{so realistic}'' that it occasionally caused misjudgments during movement. For example, he attempted to sit down on a virtual chair that had been repositioned.
P2 observed that when deeply focused on arranging virtual objects, he sometimes ``\textit{bumped into real furniture}.''
These observations highlight a tension inherent in high-fidelity MR: the more convincing the virtual content, the greater the risk that users confuse it with physical objects, or conversely, lose awareness of the physical environment.

\subsubsection{Suggestions for Improvement}
Participants offered constructive suggestions for extending \SysName{}.
Three participants (P5, P6, P8) independently requested a \emph{multi-select} capability, with P5 and P8 further suggesting a \emph{copy-and-paste} mechanism to duplicate furniture groups, which they believed would ``\textit{improve operational efficiency}.''
P6 wanted multi-select specifically to ``\textit{simultaneously adjust the size of a group of furniture}.''
P9 raised a broader point about the language-based interaction, noting that ``\textit{language alone sometimes cannot accurately describe what I want}'' and expressing a desire for the system to better infer user intent.
She gave a concrete example: when a user places a table, the system could automatically arrange nearby idle chairs around it, rather than requiring an explicit verbal command.

%% file: text/7_conclusion.tex
\section{Conclusion, Limitations, and Future Work}

This paper presents \SysName{}, a mixed-reality system for authoring real-room layouts in situ. \SysName{} presents the evolving proposal at true scale through a virtual room proxy spatially registered to the target room, while a shared layout state allow direct manipulation, language-guided editing, and intent locking to contribute to the same developing design. To support this authoring process with real-room layout updates, a room-layout design skill guides a general-purpose LLM agent through functional grouping, anchor furniture selection, and geometry-guided refinement, grounded in an axis-normalized representation of the scanned room and reusable geometric checks.

Nevertheless, several limitations remain in the current implementation. A layout generation round currently takes around 3 mins on average, limiting how rapidly users can move between evaluating a proposal and receiving a revised layout. Future work could improve efficiency by focusing agent execution on the parts of the scene involved in each modification and by using faster foundation models. The virtual room proxy also requires a careful balance between visual realism and clear virtual--physical differentiation. Adaptive visual cues could help users distinguish proposed furniture from physical objects during movement and manipulation while preserving the proxy's support for in situ comparison.

More broadly, \SysName{} illustrates a shift from generating spatial designs in isolation to authoring them with their intended physical environment in the loop. By keeping proposals spatially registered throughout generation, inspection, and revision, in situ authoring turns physical context from a final validation site into an active part of the design process.

%% file: text/supp.tex
\input{text/supp_01_summary}

\input{text/supp_03_evaluation}
\input{text/supp_04_prompts}

%% file: text/supp_01_summary.tex
\section{Summary of Supplementary Materials}

This supplementary document provides additional evaluation and user-study materials that complement the main paper:

\begin{itemize}[leftmargin=*, nosep]
    \item \textbf{Additional details on technical evaluation:} definitions of the six evaluation metrics, aggregate and per-room-type results on the 43-room benchmark, details of the LayoutVLM-PolyRepair adaptation, a paired analysis of skill-guided generation latency, and the qualitative layouts used in the perceptual evaluation.
    \item \textbf{Additional details on system usability study:} representative user-operation workflows, representative final layouts produced under the \SysName{} (MR) and \BaselineName{} (VR) conditions, and detailed questionnaire results.
    \item \textbf{Skill-guided agent instructions:} the room-layout design skill, including its referenced layout principles and output-format instructions.
\end{itemize}

%% file: text/supp_03_evaluation.tex
\section{Additional Quantitative Evaluation Details}
\label{sec:supp_quantitative_analysis}

\subsection{Metric Definitions}

We report six metrics drawn from the evaluation protocols of LayoutVLM~\cite{sun2025layoutvlm} and GLTScene~\cite{li2024gltscene}, covering physical plausibility, semantic coherence, and wall-alignment quality.

\emph{Collision-Free Score (\textbf{CF}~$\uparrow$)} measures the percentage of placed furniture items whose oriented bounding boxes do not intersect with any other item in the scene.

\emph{In-Boundary Score (\textbf{IB}~$\uparrow$)} measures the percentage of placed furniture items that lie entirely within the room boundary.

\emph{Positional Coherency (\textbf{PC}~$\uparrow$)} evaluates whether semantically related furniture items are placed near each other in a way that supports joint use. Following LayoutVLM~\cite{sun2025layoutvlm}, we render top-down and side views of each layout and prompt an LLM to rate positional coherence on a 0--100 scale.

\emph{Rotational Coherency (\textbf{RC}~$\uparrow$)} evaluates whether semantically related furniture items are oriented relative to each other in functionally appropriate ways. As with PC, we obtain this score from LLM judgments over the same top-down and side-view renderings on a 0--100 scale.

\emph{Physically-Grounded Semantic Alignment Score (\textbf{PSA}~$\uparrow$)} combines semantic coherence with physical plausibility. For each layout, an LLM rates its match to the room-type layout criteria on a 0--100 scale, and we compute $\mathrm{PSA}=\mathrm{CF}\times\mathrm{IB}\times\mathrm{LayoutMatch}$. PSA therefore penalizes layouts that are semantically plausible but physically infeasible.

We use Gemini~3~Flash~\cite{google2025gemini3flash} as the evaluator for PC, RC, and the layout-criteria-match rating used by PSA throughout.

\emph{Alignment Error (\textbf{AE}~$\downarrow$)}~\cite{li2024gltscene} quantifies how well each furniture item is oriented relative to its nearest wall:
\begin{equation}
    \mathrm{AE}(F_i) =
    \begin{cases}
        1 - \cos^{2}(2\theta), & \text{if } F_i \text{ is inside the room} \\
        1, & \text{otherwise}
    \end{cases}
    \label{eq:alignment_error}
\end{equation}
where $\theta$ is the angle between the orientation of $F_i$ and its nearest wall segment. The scene-level score is averaged over all furniture items. A value of~0 indicates perfect wall alignment, while items outside the room receive the maximum penalty of~1.

\subsection{LayoutVLM Polygonal-Boundary Adaptation}

To construct LayoutVLM-PolyRepair, we reconstruct each floor polygon from its ordered wall sequence by joining adjacent endpoints into a validated closed loop. We transform the polygon into LayoutVLM's local task frame using the minimum-area enclosing rectangle, with its longer edge defining the local $x$-axis and the orthogonal direction defining $y$. Duplicate endpoints are removed, the vertices are normalized to counterclockwise order, and the true polygon is supplied to LayoutVLM's wall representation, scene renderer, and optimizer.

During optimization, we construct the complete two-dimensional oriented bounding box $B_i$ of each furniture item $i$ from its position, yaw, and asset dimensions. Given the room polygon $P$ and an outward numerical tolerance of $\epsilon=5\,\mathrm{mm}$, we require $P_{\epsilon}$ to cover $B_i$.
The optimizer applies this constraint to every furniture item. When $B_i$ extends beyond $P_{\epsilon}$, it translates the furniture center to the nearest feasible position while keeping its yaw and dimensions fixed. The search samples candidate centers, selects the closest one whose full footprint lies inside the polygon, and refines the translation toward the current position. Using the complete footprint detects both ordinary boundary excursions and furniture edges that cross concave cut-outs.

We apply the full-footprint projection while all furniture positions are optimized together under the collision and spatial-relation objectives. The projection is repeated after positional constraints that can change furniture centers, followed by a final projection before the layout is returned.

\subsection{Skill-Guidance Generation Efficiency}

We isolate the generation-efficiency effect of the room-layout design skill in a paired experiment on 43 3D-FRONT rooms. Both conditions receive the same room and furniture inputs, the same task request, and GPT-5.6 Sol~\cite{openai2026gpt56sol} with high reasoning effort. The skill-guided condition additionally receives the complete room-layout design skill and render checker, whereas the unguided general-purpose agent receives only the task input and final-output format. Each condition produces one completed layout per room. We measure end-to-end agent execution time from the recorded trial duration and aggregate input and output tokens from the recorded model usage.

\begin{table}[t]
    \centering
    \scriptsize
    \caption{Generation efficiency on the 43-room benchmark. Time, token, and cost entries are per-room mean/median; $\Delta$ is computed from the means relative to the plain agent. Lower is better.}
    \label{tab:skill_guidance_efficiency}
    \setlength{\tabcolsep}{3pt}
    \resizebox{\columnwidth}{!}{%
    \begin{tabular}{@{}lcccc@{}}
    \toprule
    Condition & Time (s)~$\downarrow$ & Input Tokens~$\downarrow$ & Output Tokens~$\downarrow$ & Cost (\$)~$\downarrow$ \\
    \midrule
    Skill-guided & 170.29/157.23 & 208,499/193,521 & 5,397/5,232 & \$0.334/\$0.319 \\
    Plain & 327.93/237.61 & 207,449/182,042 & 9,552/8,682 & \$0.410/\$0.379 \\
    $\Delta$ vs. plain & $-48.07\%$ & $+0.51\%$ & $-43.50\%$ & $-18.47\%$ \\
    \bottomrule
    \end{tabular}
    }
\end{table}

The skill-guided agent completes a room in 170.29 seconds on average (median 157.23), compared with 327.93 seconds (median 237.61) for the plain agent. This is a 48.07\% reduction in mean generation latency, corresponding to a 1.93$\times$ speedup.

As shown in Table~\ref{tab:skill_guidance_efficiency}, the skill-guided agent uses 0.51\% more input tokens (208,499 vs.\ 207,449) but 43.50\% fewer output tokens (5,397 vs.\ 9,552) on average. According to GPT-5.6 Sol's cost rates~\cite{openai2026gpt56sol}, the estimated cost is \$0.334 versus \$0.410 per room, an 18.47\% reduction. Together with its lower latency, this demonstrates higher generation efficiency than the plain agent.

\subsection{Per-Room-Type Quantitative Analysis}

All quantitative results in this section are computed on the same 43-room benchmark used in the main paper (16 bedrooms, 13 living rooms, and 14 libraries). Table~\ref{tab:quantitative_per_room} additionally reports LayoutVLM-Original under the method's original rectangular-room input setting. Gemini~3~Flash evaluates PC, RC, and the layout-criteria-match ratings used by PSA for every method.

\begin{table}[t]
    \centering
    \small
    \caption{Per-room-type quantitative comparison of layout generation quality on non-rectangular rooms. PC, RC, and the layout-criteria-match ratings used by PSA are evaluated by Gemini~3~Flash. LayoutVLM-Original uses rectangular room input and true-polygon evaluation, whereas LayoutVLM-PolyRepair uses the true polygon with full-footprint boundary-constrained optimization. Best results are in \textbf{bold}; second-best are \underline{underlined}.}
    \label{tab:quantitative_per_room}
    \setlength{\tabcolsep}{3.5pt}
    \begin{tabular}{l cccccc}
    \toprule
    Method & CF~$\uparrow$ & IB~$\uparrow$ & PC~$\uparrow$ & RC~$\uparrow$ & PSA~$\uparrow$ & AE~$\downarrow$ \\
    \midrule
    \multicolumn{7}{l}{\textit{Bedroom} ($n{=}16$)} \\
    \midrule
    GLTScene~\cite{li2024gltscene}       & \underline{0.8927}   & \textbf{1.0000}   & \underline{56.12}   & 52.19   & \underline{47.86}   & \underline{0.0008}   \\
    LayoutVLM-Original~\cite{sun2025layoutvlm}    & 0.6167   & 0.5188   & 43.75   & 35.88   & 14.88   & 0.2261   \\
    LayoutVLM-PolyRepair~\cite{sun2025layoutvlm}   & 0.8604   & 0.5437   & 55.69   & \underline{53.38}   & 28.14   & 0.0569   \\
    \SysName{}                            & \textbf{1.0000}   & \textbf{1.0000}   & \textbf{63.62}   & \textbf{59.94}   & \textbf{59.88}   & \textbf{0.0000}   \\
    \midrule
    \multicolumn{7}{l}{\textit{Living Room} ($n{=}13$)} \\
    \midrule
    GLTScene~\cite{li2024gltscene}       & 0.6515   & \underline{0.9716}   & 50.15   & 48.69   & 30.73   & \underline{0.1067}   \\
    LayoutVLM-Original~\cite{sun2025layoutvlm}    & 0.6844   & 0.4749   & 44.15   & 38.00   & 14.56   & 0.5153   \\
    LayoutVLM-PolyRepair~\cite{sun2025layoutvlm}   & \underline{0.9013}   & 0.8677   & \underline{55.08}   & \underline{57.00}   & \underline{44.67}   & 0.1292   \\
    \SysName{}                            & \textbf{0.9330}   & \textbf{1.0000}   & \textbf{57.69}   & \textbf{58.62}   & \textbf{54.09}   & \textbf{0.0273}   \\
    \midrule
    \multicolumn{7}{l}{\textit{Library} ($n{=}14$)} \\
    \midrule
    GLTScene~\cite{li2024gltscene}       & \underline{0.8549}   & \underline{0.8826}   & 57.36   & 49.57   & \underline{38.18}   & 0.1406   \\
    LayoutVLM-Original~\cite{sun2025layoutvlm}    & 0.2992   & 0.6643   & 49.43   & 42.79   & 10.30   & 0.2740   \\
    LayoutVLM-PolyRepair~\cite{sun2025layoutvlm}   & 0.8051   & 0.8135   & \textbf{77.36}   & \underline{54.14}   & 37.85   & \underline{0.0425}   \\
    \SysName{}                            & \textbf{0.9796}   & \textbf{1.0000}   & \underline{75.21}   & \textbf{55.57}   & \textbf{52.67}   & \textbf{0.0001}   \\
    \bottomrule
    \end{tabular}
\end{table}

The room-type breakdown shows that \SysName{} leads all six metrics in living rooms, with CF~0.9330, IB~1.0000, PC~57.69, RC~58.62, PSA~54.09, and AE~0.0273. In bedrooms, \SysName{} likewise leads CF, PC, RC, PSA, and AE and ties GLTScene at IB~1.0000.

Across the 14 library rooms, LayoutVLM-PolyRepair obtains the highest PC (77.36), while \SysName{} is second on PC (75.21) and leads CF (0.9796), IB (1.0000), RC (55.57), PSA (52.67), and AE (0.0001). GLTScene ranks second on library CF (0.8549), IB (0.8826), and PSA (38.18). LayoutVLM-Original's rectangular-room setting records lower physical and semantic scores under true-polygon evaluation across all three room categories. Thus, \SysName{} maintains the strongest overall metric profile in every room category while LayoutVLM-PolyRepair is especially competitive on library positional coherence.

\section{Qualitative Results used in Perceptual Evaluation}

\begin{figure*}[t]
    \centering
    \includegraphics[width=\textwidth]{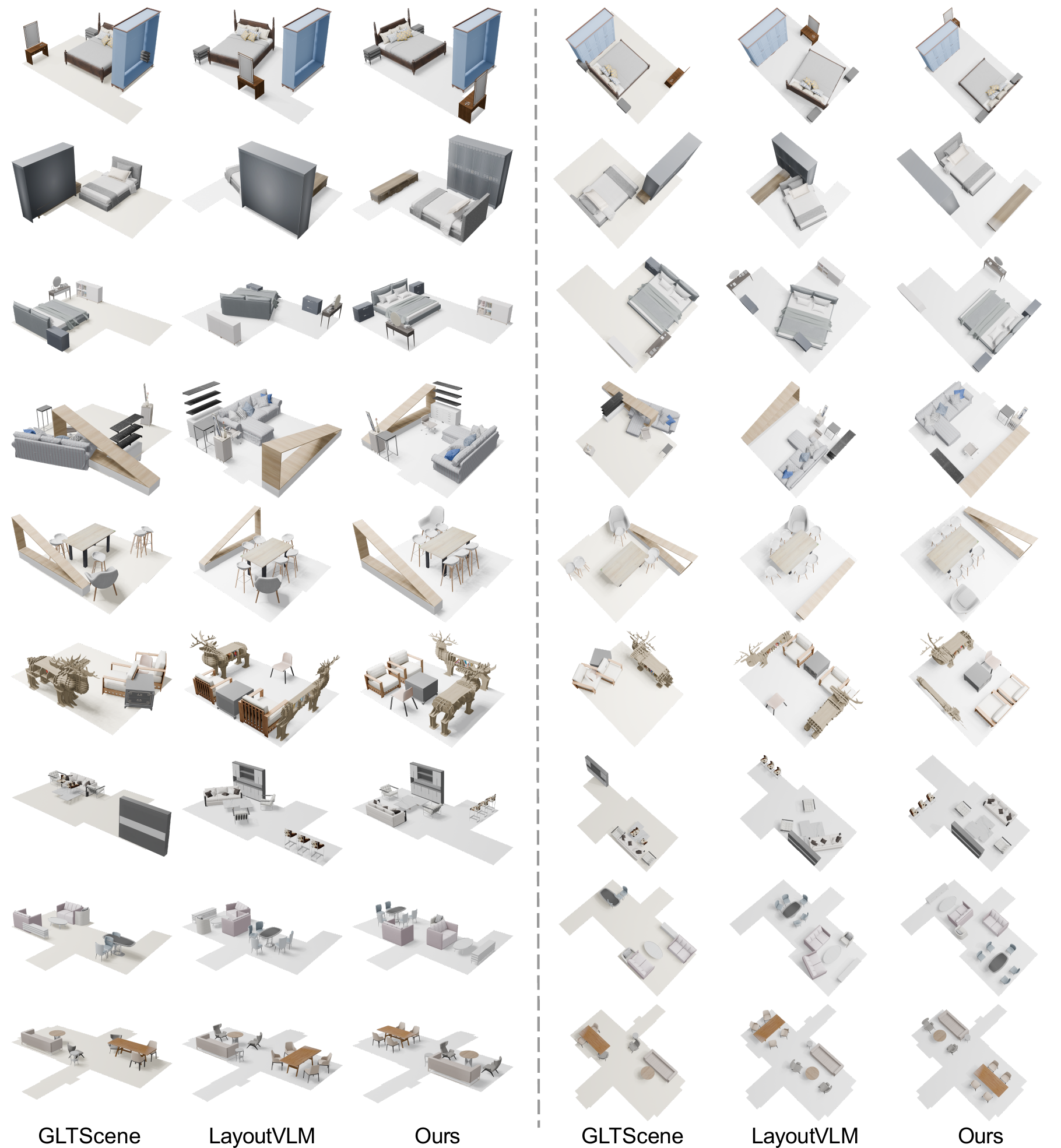}
    \caption{Qualitative layouts used in the perceptual evaluation, shown from perspective (left) and top-down (right) views. Within each view, the columns show GLTScene, LayoutVLM-PolyRepair, and \SysName{} from left to right.}
    \label{fig:supp_perceptual_evaluation}
\end{figure*}

In this section, we report the qualitative results used in the perceptual evaluation comparison between \SysName{} and its baselines, GLTScene and LayoutVLM-PolyRepair, in Figure~\ref{fig:supp_perceptual_evaluation}.

\section{Representative Workflows}
\label{sec:supp_task_env}

\begin{figure*}[t]
    \centering
    \includegraphics[width=\textwidth]{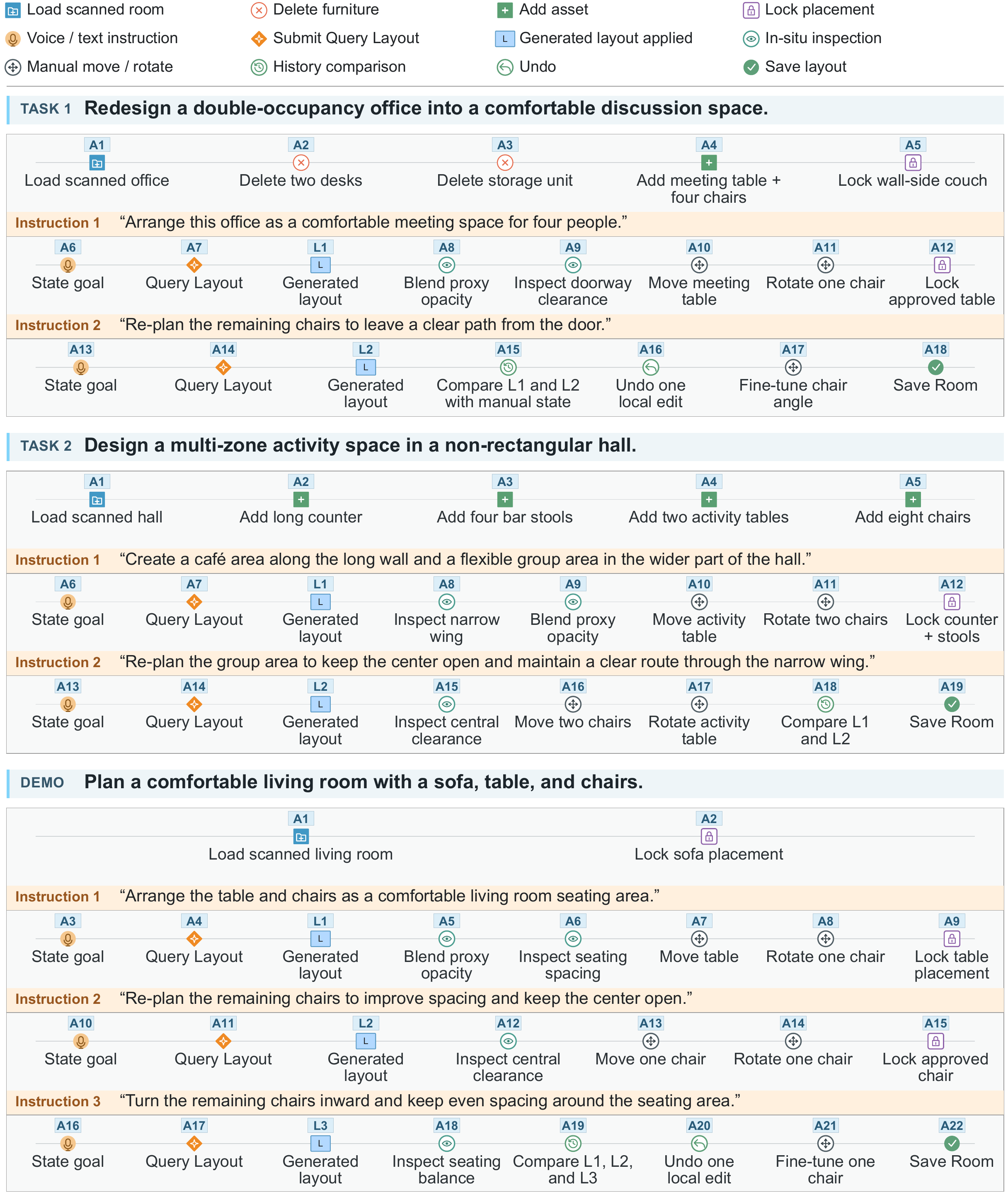}
    \caption{Illustrative user-operation sequences for the two system-usability-study tasks and one supplementary living-room demonstration scenario. Action nodes ($A_i$) denote user operations, and layout nodes ($L_i$) denote system-generated layouts. Each sequence shows room and asset preparation, an initial language-guided generation round, in situ inspection and manual refinement, intent locking, one or more follow-up generation rounds driven by refined instructions, and final comparison or adjustment before the room is saved.}
    \label{fig:user_workflow}
\end{figure*}

Figure~\ref{fig:user_workflow} presents 3 illustrative user-operation sequences, which include two assigned study tasks and a supplementary living-room demonstration scenario described in the demo materials.

\section{Layout Results from the System Usability Study}
\label{sec:supp_user_study_results}

\begin{figure*}[t]
    \centering
    \includegraphics[width=\textwidth]{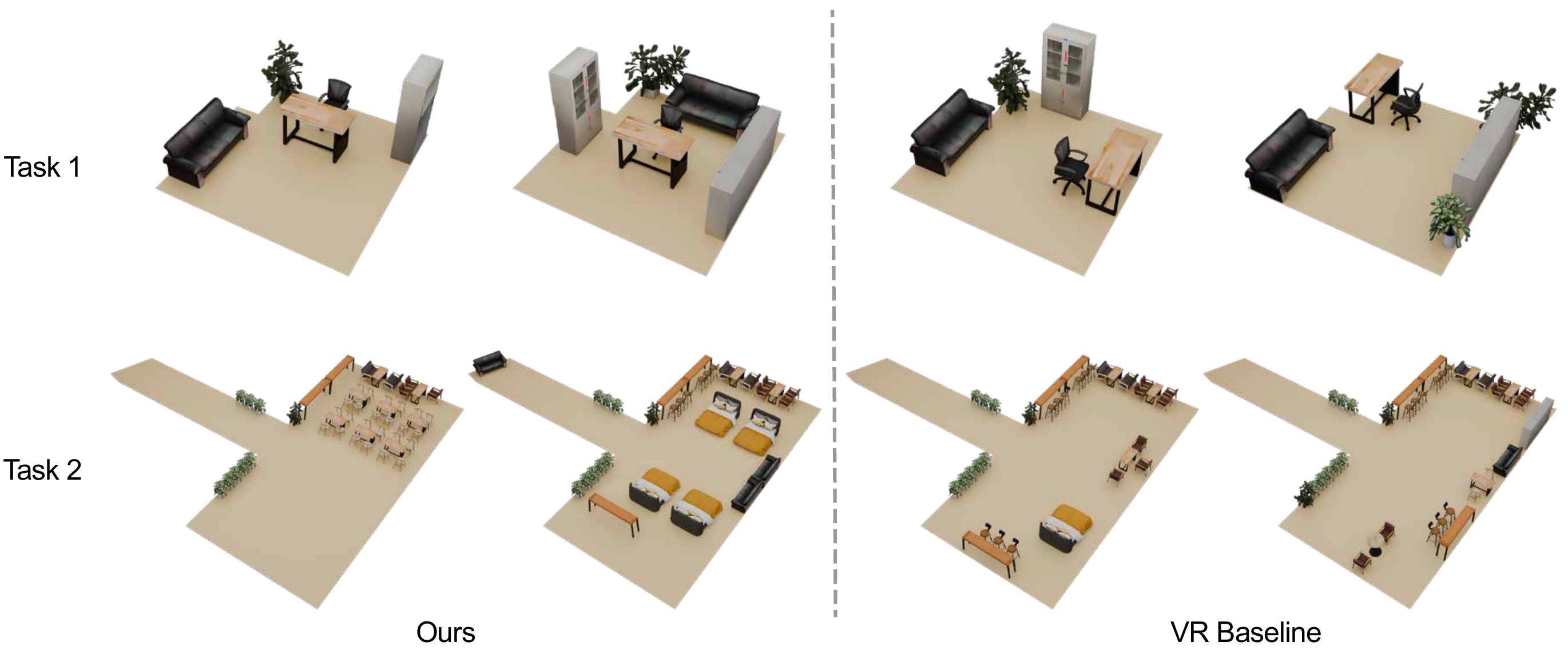}
    \caption{Selected layout results from participants in the system usability study. Each row corresponds to one task (Task~1: office redesign; Task~2: open hall planning).}
    \label{fig:user_study_result}
\end{figure*}

Figure~\ref{fig:user_study_result} presents selected layout results from the system usability study described in user study. For each task, we show perspective views of final layouts produced under the \SysName{} (MR, in situ) and \BaselineName{} (VR, off-site) conditions, chosen from across all participants to illustrate representative outcomes.

Across both tasks, the final layouts produced under the two conditions are qualitatively comparable: furniture groupings form coherent functional zones, items are properly placed within the room boundary, and the overall spatial organization reflects reasonable design intent. This outcome is expected, because the two conditions share the same underlying skill-guided layout agent and the same set of interaction modalities (i.e. direct manipulation, language-guided editing, and intent locking). Moreover, participants in both conditions were free to iteratively refine the agent-generated proposals until they were satisfied with the result. Given identical algorithmic support and unconstrained editing time, users naturally converge on layouts of similar quality regardless of whether the planning environment is MR or VR.

This observation reinforces the interpretation of the questionnaire and interview findings reported in the main paper (user study section): the primary advantage of in situ mixed-reality planning does not manifest as measurably superior final layouts, but rather as an improved planning experience. Participants using \SysName{} reported significantly better spatial perception, higher layout confidence, and lower task load. Significant reductions occurred in temporal demand, effort, performance, and frustration. Final designs remained comparable in quality to those produced in VR. In situ MR grounding enables users to reach equivalent design outcomes with greater confidence, because the physical room context reduces the cognitive effort required to evaluate spatial relationships with architectural features such as doorways and walls.

%% file: text/supp_04_prompts.tex
\section{Skill-Guided Agent Instructions}

Each layout generation begins with the same instruction: ``Read \texttt{skill/SKILL.md} completely and follow it to complete this room-layout trial.'' The session contains the task-specific scene evidence and user request, together with an output schema that represents each movable furniture item by its position and orientation in continuous coordinates.

The room-layout design skill guides the agent in developing a layout from this evidence. The skill and its two referenced instruction files are reproduced below.

\onecolumn
\begin{tcblisting}{
    colback=gray!10,
    colframe=gray!50,
    listing only,
    breakable,
    width=\textwidth,
    listing options={basicstyle=\ttfamily\small, breaklines=true, columns=fullflexible, keepspaces=true, showstringspaces=false, showlines=true},
}
---
name: arrange-room
description: Use Codex to create, render, review, and revise continuous movable-furniture layouts around optional fixed furniture from top-down images plus JSON data. Use for prepared trials that require continuous X/Z positions, front yaw, geometry feedback, and final layout selection.
---

# Arrange a Room

Codex chooses every furniture position, orientation, and revision from the supplied evidence. The included script renders each complete layout and writes geometry feedback.

## Use the evidence

Work from the prepared trial's `input/` directory:
- `room.png` and `room.json` show the same room in X/Z coordinates, including its boundary, openings, any initial movable layout, and fixed furniture. Initial movable furniture is green and labeled `[initial]`; fixed furniture is gray, hatched, and labeled `[fixed]`.
- `furniture.png` and `furniture.json` show the movable catalog, exact footprint dimensions, IDs, categories, and front arrows.
- `request.md` contains room-specific requirements.
- `layout.schema.json` describes the continuous output format.

References for guidance:
- Use the files under `input/` as the design evidence
- Read [layout principles](references/layout-principles.md) before planning the layout
- Read [layout format](references/layout-format.md) before writing JSON.

## Review and revise

Write one complete layout with a footprint center and front yaw for every movable furniture item. Do not include fixed furniture in `layout.json` or change its supplied pose. Render when feedback would help. Inspect `layout.png` and `feedback.json`. Keep the layout, adjust selected items, revise the overall arrangement, or finish based on that evidence.

Choose the size and timing of each revision from the evidence. A `finalizable` value of `true` confirms the inventory, boundary, and overlap checks. Use visual review to assess circulation, access, functional groups, and balance. Continue until the layout passes these checks and has a clear arrangement, then promote the chosen attempt with `--promote`.

After final selection succeeds, inspect `final/feedback.json` and `final/manifest.json`, report the result, and stop.

Each render call writes a new attempt directory. Promoting an attempt does not render it again. Revise `layout.json` when another layout is useful. Use `needs_review` when the input files are incomplete or inconsistent, or when every complete layout violates geometry, circulation, or access requirements.

## Acceptance criteria

Before final selection, confirm that:

- Every movable furniture item appears exactly once, and no fixed furniture appears in the placements.
- Every boundary violation and unintended movable-to-movable or movable-to-fixed overlap is resolved.
- Doors, main routes, furniture fronts, and access clearances remain usable.
- Each orientation is consistent with the furniture's front-direction arrow.
\end{tcblisting}
\refstepcounter{figure}
\label{fig:skill_instructions}
\noindent\begin{minipage}{\textwidth}
\centering
\textbf{Figure~\thefigure:} Room-layout design skill provided to the agent.
\end{minipage}
\vspace{\baselineskip}

\begin{tcblisting}{
    colback=gray!10,
    colframe=gray!50,
    listing only,
    breakable,
    width=\textwidth,
    listing options={basicstyle=\ttfamily\small, breaklines=true, columns=fullflexible, keepspaces=true, showstringspaces=false, showlines=true},
}
# Layout principles

## Read image and JSON evidence

- Cross-check the room image against `room.json`; use the image for shape and circulation, and JSON for precise coordinates and opening segments.
- Cross-check the furniture sheet against `furniture.json`; use the footprint, dimensions, category, ID, and red front-direction arrow together.
- Treat hatched `[fixed]` furniture in the room as immutable occupied space. Preserve access to it and arrange only the movable catalog around it.
- Treat green `[initial]` furniture as the current movable layout, not as a constraint. Preserve useful parts of it when the request needs only a local change.
- Use the light metre reference lines to estimate scale and continuous coordinates.
- Reserve about 0.5 m for primary circulation and front access when the room permits it.

## Plan the layout

- Functional Grouping: organize the room at both the zone and furniture scales.
  - Functional zoning: establish coherent activity areas such as reading, working, dining, and relaxation zones.
  - Furniture grouping: arrange related furniture within each zone according to functional adjacency and orientation, such as nightstands beside a bed, seating around a table, a coffee table in front of a sofa, and media storage opposite seating.
- Anchor Furniture Selection: identify the large or function-defining furniture that anchors each zone, place it first according to room boundaries and functional orientation, and arrange related items around it.
- Geometry-Guided Refinement: use image and JSON geometry, fixed obstacles, access clearances, rendered layouts, and geometry feedback to detect conflicts and iteratively refine positions and orientations.
- Prioritize access and circulation over symmetry: float tables and selected seating when this improves access or grouping, and keep doorways, entry paths, inter-zone routes, and storage fronts clear before optimizing symmetry.

## Review and revise

- Match the revision scope to the issue: make targeted adjustments for isolated problems, and reconsider the overall arrangement when multiple items compete for space.
- Review function and composition together: use each rendered layout to assess orientation, access, circulation, functional relationships, and visual balance.
\end{tcblisting}
\refstepcounter{figure}
\label{fig:skill_layout_principles}
\noindent\begin{minipage}{\textwidth}
\centering
\textbf{Figure~\thefigure:} Layout principles referenced by the room-layout design skill.
\end{minipage}
\vspace{\baselineskip}

\begin{tcblisting}{
    colback=gray!10,
    colframe=gray!50,
    listing only,
    breakable,
    width=\textwidth,
    listing options={basicstyle=\ttfamily\small, breaklines=true, columns=fullflexible, keepspaces=true, showstringspaces=false, showlines=true},
}
# Continuous layout format

## Evidence and coordinates

Use the matching visual and numeric evidence under `input/`. Coordinates are metres in the X/Z room coordinates recorded by `room.json`. Set positions directly in these coordinates; do not snap them to a grid. A placement position is the center of the furniture footprint. `front_world_yaw_deg=0` faces `+Z`, `90` faces `+X`, and positive angles turn clockwise in the supplied top-down view.

`room.json.fixed_furniture` lists immutable furniture already present in the room. It is rendered with hatching and a `[fixed]` label. Use it for collision, access, circulation, and functional grouping, but do not copy it into `layout.json`.

When present, `room.json.initial_layout` lists the current positions and front yaws of movable furniture. It is rendered in green with an `[initial]` label. Use it as the starting state, but still list every movable item in the output layout.

## Layout JSON

Write strict JSON as `layout.json` in the trial root:

```json
{
  "schema_version": 1,
  "status": "completed",
  "placements": [
    {
      "furniture_id": "KING-SIZE BED#0",
      "x": -2.263,
      "z": 3.4413,
      "front_world_yaw_deg": 180.0
    }
  ],
  "summary": "The bed group uses the quiet end of the room while preserving the entry route.",
  "accepted_overlaps": []
}
```

List every movable ID from `input/furniture.json` exactly once and list no fixed IDs. Values may use any finite precision. `x`, `z`, and `front_world_yaw_deg` define the rendered position and orientation. Keep `accepted_overlaps` empty unless feedback reports a specific pair with an intended functional overlap. An accepted item has the form `{"items": ["DINING CHAIR#0", "DINING TABLE#0"], "reason": "The chair is intentionally tucked under the table."}`.

If the input files are incomplete or inconsistent, use `status: "needs_review"`, an empty placement list, an empty accepted-overlap list, and explain the issue in `summary`.

## Render and feedback

From the trial root, render a layout with the copied Skill script:

```
uv run skill/scripts/render_layout.py --input input --layout layout.json --workspace .
```

The command creates an `attempts/aNNN/` directory containing the submitted layout, a top-down `layout.png`, and `feedback.json`. Inspect the image and report for boundary violations and overlap pairs. Each overlap reports `fixed_items` so movable-to-fixed conflicts are explicit. Revise movable placements and invoke the command again when another layout is useful.

When satisfied, promote the chosen finalizable attempt by its reported ID:

```
uv run skill/scripts/render_layout.py --workspace . --promote aNNN
```

The command does not create or render another attempt. It copies the chosen attempt's `layout.json` and `feedback.json`, copies `layout.png` when present, and writes `final/manifest.json`. A `needs_review` attempt can also be promoted. After the trial, the controller exports the scene and runs external evaluation.
\end{tcblisting}
\refstepcounter{figure}
\label{fig:skill_layout_format}
\noindent\begin{minipage}{\textwidth}
\centering
\textbf{Figure~\thefigure:} Layout-format instructions referenced by the room-layout design skill.
\end{minipage}
\vspace{\baselineskip}

\twocolumn